\documentclass[12pt]{article}
\usepackage{epsfig}
\usepackage{graphics}
\usepackage{psfrag}
\usepackage{amssymb}

\newcommand{\beq}{\begin{equation}}
\newcommand{\eeq}{\end{equation}}
\newcommand{\bea}{\begin{eqnarray}}
\newcommand{\eea}{\end{eqnarray}}

\newcommand{\e}{\mbox{e}}
\renewcommand{\d}{\mbox{d}}

\newcommand{\lam}{\lambda}
\newcommand{\La}{\Lambda}
\renewcommand{\b}{\beta}
\renewcommand{\a}{\alpha}
\newcommand{\n}{\nu}

\newcommand{\Om}{\Omega}

\newcommand{\del}{\delta}

\newcommand{\kp}{\kappa}

\newcommand{\oh}{\frac{1}{2}}

\newcommand{\dg}{\dagger}

\newcommand{\tr}{\mathrm{tr}\,}
\newcommand{\Tr}{\mathrm{Tr}\,}
\newcommand{\ra}{\rangle}
\newcommand{\rra}{\right\ra}
\newcommand{\la}{\langle}
\newcommand{\lla}{\left\la}
\newcommand{\prt}{\partial}
\newcommand{\mi}{\!-\!}

\newcommand{\pl}{\!+\!}

\newcommand{\cD}{{\cal D}}

\newcommand{\cT}{{\cal T}}

\newcommand{\cO}{{\cal O}}

\newcommand{\tG}{{\tilde{G}}}

\newcommand{\tW}{{\tilde{W}}}
\newcommand{\tw}{{\tilde{w}}}

\newcommand{\tx}{{\tilde{x}}}

\newcommand{\tH}{{\tilde{H}}}

\newcommand{\hH}{{\hat{H}}}

\newcommand{\hG}{{\hat{G}}}

\newcommand{\hx}{{\hat{x}}}

\newcommand{\hp}{{\hat{p}}}

\newcommand{\bx}{{\bar{x}}}

\newcommand{\slt}{\sqrt{\lam} t}
\newcommand{\sla}{\sqrt{\lam}}
\newcommand{\sOm}{\sqrt{\Om}}

\newcommand{\vac}{|0\ra}
\newcommand{\cav}{\la 0 |}
\newcommand{\dll}{\frac{dl}{l}}

\newcommand{\rf}[1]{(\ref{#1})}

\begin{document}

\begin{center}

{ \large \bf Proper time is stochastic time in 2d quantum gravity}\footnote{Talk presented at the meeting ``Foundations of Space and Time'', Cape Town, 
10-14 August 2009. To appear in the proceedings, CUP}

\vspace{30pt}

{\sl J.\ Ambj\o rn}$\,^{a,b}$, {\sl R.\ Loll}$\,^{b}$,
{\sl Y.\ Watabiki}$\,^{c}$, {\sl W.\ Westra}$\,^{d}$ and
{\sl S.\ Zohren}$\,^{e,f}$

\vspace{24pt}

{\footnotesize
$^a$~The Niels Bohr Institute, Copenhagen University\\
Blegdamsvej 17, DK-2100 Copenhagen \O , Denmark.\\
{ email: ambjorn@nbi.dk}\\

\vspace{10pt}

$^b$~Institute for Theoretical Physics, Utrecht University, \\
Leuvenlaan 4, NL-3584 CE Utrecht, The Netherlands.\\
{ email: loll@phys.uu.nl}\\

\vspace{10pt}

$^c$~Tokyo Institute of Technology,\\ 
Dept. of Physics, High Energy Theory Group,\\ 
2-12-1 Oh-okayama, Meguro-ku, Tokyo 152-8551, Japan\\
{email: watabiki@th.phys.titech.ac.jp}\\

\vspace{10pt}

$^d$~Department of Physics, University of Iceland,\\
Dunhaga 3, 107 Reykjavik, Iceland\\
{ email: wwestra@raunvis.hi.is}\\

\vspace{10pt}

$^e$~Mathematical Institute, Leiden University,\\
Niels Bohrweg 1, 2333 CA Leiden, The Netherlands\\
{email: zohren@math.leidenuniv.nl}\\

\vspace{10pt}

$^f$~Department of Statistics, Sao Paulo University,\\
Rua do Matao, 1010, 05508-090, Sao Paulo, Brazil

}
\vspace{24pt}

\end{center}

\begin{center}
{\bf Abstract}
\end{center}

We show that proper time, when defined in the quantum theory 
of 2d gravity, becomes identical to the stochastic time associated
with the stochastic quantization of space. This observation 
was first made by Kawai and collaborators in the context of 2d Euclidean
quantum gravity, but the relation is even simpler and more transparent 
in he context of 2d gravity formulated in the framework of CDT (causal 
dynamical triangulations).
%

\newpage

\section{introduction}\label{intro}

 Since time plays such a prominant 
role in ordinary flat space  quantum field
theory defined by a Hamiltonian, it is of interest to study the 
role of time even in toy models of quantum gravity where 
the role of time is much more enigmatic.
The model we will describe in this article is the so-called
causal dynamical triangulation (CDT) model of quantum gravity.
It starts by providing an
utraviolet regularization in the form of a lattice theory, 
the lattice link length being the
(diffeomorphism invariant) UV cut-off. In
addition the lattice respects causality.  
It is formulated in the spirit of {\it asymptotic safty}, where it 
is assumed that quantum gravity is described entirely by 
``conventional'' quantum field theory, in this case by
approaching a non-trivial fixed point
\cite{weinberg}, \cite{reuteretc} . 
It is formulated in space-times with Lorentzian signature, 
but the regularized space-times which are used 
in the path integral defining the theory  allow
a rotation to Euclidean space-time. The action used is the 
Regge action for the piecewise linear geometry. Each (piecewise linear)
geometry used in the path integral has after rotation
to Euclidean signature an Euclidean Regge action, related to the 
original Lorentzian action in the same way as when one in flat space-time  
rotates Lorentzian time to Euclidean time (see \cite{ajl4d,blp} for details
of the Regge action in the CDT approach). By rotating each Lorentzian 
CDT lattice (or piecewise linear geometry) to Euclidean signature 
the non-perturbative path integral is performed
by summing over a set of Euclidean lattices originating from the Lorentzian lattices
with a causal stucture, and this set is different from the full set of piecewise linear 
Euclidean lattices.  Like in ordinary lattice field theories
we approach the continuum by fine-tuning the bare coupling constants.
The rotation to Euclidean space-time makes it possible to use  Monte Carlo 
simulations when studying the theory, and in four-dimensional space-time, which 
for obvious reasons has our main interest, there exists a region
of coupling constant space  where
the infrared behavior of the universe seen by the computer is that 
of (Euclidean) de Sitter space-time \cite{agjl,emergence} (for  
a pedagodical review, see \cite{causality}).  
Non-trivial UV properties have been observed 
\cite{semiclassical},
properties which have been reproduced by other ``field theoretical'' 
approaches to quantum gravity \cite{rg1,horava}.

Numerical simulations are very useful when trying to understand if a 
non-perturbatively defined quantum field theory has a chance
to make sense.  However, numerical simulations have their
limitations in the sense that they will never provide a proof
of the existence of a theory and it might be difficult in detail
to follow the way the continuum limit is approached since 
it requires larger and larger lattices. It is thus of interest
and importance to be able to study this in detail, even if
only in a toy model. Two-dimensional quantum gravity is such 
a toy model which has a surprisingly rich structure. Many of the 
intriguing questions in quantum gravity and in lattice quantum gravity
are still present in the two-dimensional theory.
We will discuss the solution to two-dimensional CDT in the rest 
of this article and we will see that in  {\it time}, which is introduced
as {\it proper time}, has an
interpretation as {\it stochastic time} in a process there the
evolution of space can be viewed as a stochastic.

\section{The CDT formalism}\label{cdt}

We start from Lorentzian simplicial 
spacetimes with $d=2$ and insist
that only causally well-behaved geometries appear in the
(regularized) Lorentzian path integral. 
A crucial property of our explicit construction is that 
each of the configurations allows for a rotation to Euclidean signature,
as mentioned above.
We rotate to a Euclidean regime in order to perform the sum over geometries
(and rotate back again afterward if needed).
We stress here (again) that although the sum is performed over geometries 
with Euclidean signature, it is different from what one would 
obtain in a theory of quantum gravity based {\it ab initio} on Euclidean spacetimes.
The reason is that not all Euclidean geometries with a given topology are 
included in the ``causal'' sum since in general they have no correspondence 
to a causal Lorentzian geometry.
 
\begin{figure}[t]

\centerline{\scalebox{0.3}{\rotatebox{0}{\includegraphics{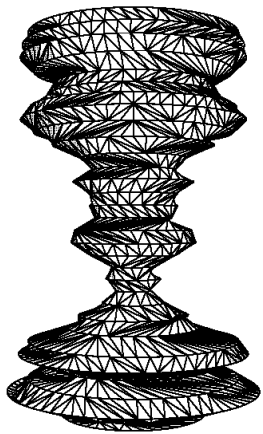}}}}
\caption{Piecewise linear space-time histories 1+1 dimensional quantum gravity}
\label{fig0}
\end{figure}

\begin{figure}[t]
\centerline{\scalebox{0.6}{\rotatebox{0}{\includegraphics{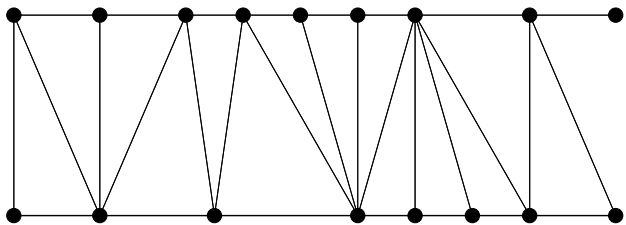}}}}
\caption{The propagation of a spatial slice from time $t$ to time $t+1$.
The end of the strip should be joined to form a band with topology $S^1\times
[0,1]$.}
\label{2dminkowski}
\end{figure}

We refer to \cite{ajl4d} for a detailed description
of how to construct the class of piecewise linear geometries used 
in the Lorentzian path integral in higher dimensions. 
The most important assumption is the existence of 
a global proper-time foliation. This is illustrated
in Fig.\ \ref{fig0} in the case of two dimensions. We have a sum over 
two-geometries, ``stretching'' between two ``one-geometries'' separated 
a proper time $t$ and constructed from two-dimensional building blocks.
In Fig.\ \ref{2dminkowski} we have shown how to fill the two-dimensional
space-time between the space (with topology $S^1$) at time $t_n$ and 
time $t_{n+1}=t_n+a$ where $a$ denotes the lattice spacing. While we in the 
lattice model often use units where everything is measured 
in lattice length (i.e.\ the lattice links have length one),
we are of course interested in taking the limit $a \to 0$ to recover 
continuum physics.

In the path integral we will be summing over all possible ways 
to connect a given 1d ``triangulation'' at time $t_n$ and a given
1d triangulation at $t_{n+1}$ to a slab of 2d space-time as shown 
in Fig.\ \ref{2dminkowski},  
and in addition we will sum over all 1d ``triangulations''
of $S^1$ at times $t_n$. Thus we are demanding that the 
time-slicing is such that the topology of space does not change
when space ``evolves'' from time $t_n$ to time $t_{n+1}$. 

The Einstein-Hilbert action $S^{\rm EH}$ in two dimensions is 
trivial since there is no curvature term as
long as the topology of space-time is unchanged (which we 
assume presently).  
Thus the (Euclidean) action simply consist of the cosmological term:  
\beq
S_E^{\rm EH}= \lam\int \d^2 x \;\sqrt{g}~~
\longrightarrow~~
 S_E^{\rm Regge}= 
\La N_2
\label{actshort}
\eeq 
where $N_2$ denotes the total number of triangles in
the two-dimensional triangulation. We denote the discretized
action {\it the Regge action} since it is a trivial example
of the natural action for piecewise linear geometries introduced 
by Regge \cite{regge}. The dimensionless lattice cosmological coupling
constant $\La$ will be related to the continuum cosmological 
coupling constant $\lam$ by an additive renormalization:
\beq\label{renor}
\La = \La_0 + \oh \lam\, a^2,
\eeq
the factor 1/2 being conventional.
The path integral or partition function for the CDT version of 
quantum gravity is now
\bea\label{2.1}
G^{(0)}_\lam(l_1,l_2;t) &=& \int \cD [g] \; 
\e^{-S_E^{\rm EH}[g]} ~~~\to \nonumber\\ 
G^{(0)}_\La(L_1,L_2,T) &=& 
\sum_{\cT} \frac{1}{C_\cT} \; \e^{-S_E^{Regge}(\cT)},
\eea
where the summation is over all causal triangulations $\cT$ of the kind 
described above with a total of $T$ time steps, an ``entrance loop'' 
of length $l_1 = L_1 a$ and an ``exit loop'' of length $l_2= L_2 a$. 
The factor $1/C_\cT$ is a symmetry factor, given by the order of 
the automorphism group of the triangulation $\cT$.

One can, somewhat surprisingly, evaluate 
the sum over triangulations in \rf{2.1} 
analytically \cite{al}. It is a counting problem
and thus we introduce the corresponding  generating
function. In our model the generating function has a 
direct physical interpretation. We define
\beq\label{genfun}
\tG^{(0)}_\La(X_1,X_2;t) = \sum_{L_1,L_2} 
\e^{-X_1L_1}\e^{-X_2L_2} G^{(0)}_\La(L_1,L_2;T).
\eeq
Thus $\tG^{(0)}_\La(X_1,X_2;T)$ is the generating function of the 
numbers $G_\La^{(0)}(L_1,L_2;T)$ if we write $Z_1 = e^{-X_1}$, $Z_2=e^{-X_2}$.
But we can also view $X$ as a (bare) dimensionless boundary cosmological
constant, such that a boundary cosmological term $X \cdot L$ has 
been added to the action. In this way $G^{(0)}_\La(X_1,X_2;T)$ represents
the sum over triangulations where the lengths of the boundaries are 
allowed to fluctuate, the fluctuations controlled by the values  $X_i$ 
of the boundary cosmological constants.  In general we expect, just
based on standard dimensional analysis, the boundary cosmological 
constants $X_i$ to be subjected to  additive renormalizations when 
the continuum limit is approached. Like \rf{renor} we expect
\beq\label{renor1}
X_i = X_c +x_i\, a,
\eeq
where $x$ then denotes the continuum boundary cosmological constant,
and one, after renormalization, has the continuum boundary 
cosmological action $x\cdot l$.

We refer to \cite{al} for the explicit combinatorial arguments which 
allow us to find $\tG^{(0)}_\La(X_1,X_2;T)$. 
Let us just state the following results:
one can derive an exact iterative equation (using notation $Z=e^{-X}$,
$W=e^{-Y}$, $Q=e^{-\La}$)
\beq\label{cdt1}
\tG^{(0)}_\La(Z,W;T) = 
\frac{Q Z}{1- Q Z}\; \tG^{(0)}_\La\Big(\frac{Q}{1-Q Z},W;T-1\Big)
\eeq
This equation can be iterated and the solution found \cite{al}. However, it is
easy to see that $Q_c = 1/2$ and that $Z_c = 1$ and
we can now take the continuum limit in \rf{cdt1} using $t=T \cdot a$ 
and find
\beq\label{cdt32} 
\frac{\prt}{\prt t} \tG^{(0)}_\lam(x,y;t) + \frac{\prt}{\prt x}
\Bigl[ (x^2-\lam) \tG^{(0)}_\lam(x,y;t) \Bigr]=0,
\eeq
This is a standard first order partial differential equation which 
should be solved with the boundary condition 
\beq\label{cdt78}
\tG^{(0)}_\lam(x,y;t=0)=\frac{1}{x+y}
\eeq
corresponding to
\beq\label{cdt79}
G^{(0)}_\lam(l_1,l_2;t=0)=\del(l_1-l_2).
\eeq
The solution is thus 
\beq\label{cdt33}
\tG^{(0)}_\lam(x,y;t) = \frac{\bar{x}^2(t;x)-\lam}{x^2-\lam}\; 
\frac{1}{\bar{x}(t;x)+y}, 
\eeq
where $\bar{x}(t;x)$ is the solution to the characteristic equation
\beq\label{cdt34}
\frac{\d \bar{x}}{\d t} = -(\bar{x}^2-\lam),~~~~\bar{x}(t=0)=x.
\eeq
We thus have an explicit solution for $\tG^{(0)}_\lam(x,y;t) $ since we obtain
\beq\label{cdt35}
\bar{x}(t) = \sla \; 
\frac{(\sla+x)-\e^{-2\slt}(\sla-x)}{(\sla+x)+\e^{-2\slt}(\sla-x)}.
\eeq

If we interpret the propagator $G_\lam^{(0)}(l_1,l_2;t)$ as the matrix element
between two boundary states of a Hamiltonian evolution in 
``time'' $t$,
\beq\label{ham}
G^{(0)}_\lam(l_1,l_2;t)=<l_1|\e^{-{H}_0 t}|l_2>
\eeq 
we can, after an inverse Laplace transformation, read off the functional form
of the Hamiltonian operator $H_0$ from \rf{cdt32},
\beq\label{35b}
\tH_0(x) = \frac{\prt}{\prt x} \left(x^2-\lam) \right),~~~~~ {H}_0(l)=
 -l \frac{\partial^2}{\partial l^2}+\lam l .
\eeq

This end our short review of basic  2d-CDT.
We have here emphasized that all continuum results
can  be obtained by explicit solving the lattice model and 
taking the continuum limit simply by letting the lattice 
spacing $a \to 0$. The same will be true for the generalized
CDT model described below, but to make the presentation
more streamlined we will drop the explicit route via a lattice
and work directly in the continuum.

\section{Generalized CDT}\label{cap1}

It is natural the ask what happens if the strict requirement
of ``classical'' causality on each geometry appearing in
the path integral is relaxed. While causality is a resonable 
requirement as an outcome of a sensible physical theory, there
is no  compleling reason to impose it on each individual geometry
in the path integral, since these are not physical observables.
We used it, inspired by \cite{teitelboim}, as a guiding principle 
for obtaining a path integral which is different from the standard
Euclidean path integral, which was seemingly a necessity in 
higher than two space-time dimensions since a "purely"  Euclidean
higher dimensional path integral did not lead to interesting 
continuum theories.

\begin{figure}[t]
\centerline{\scalebox{0.4}{\rotatebox{0}{\includegraphics{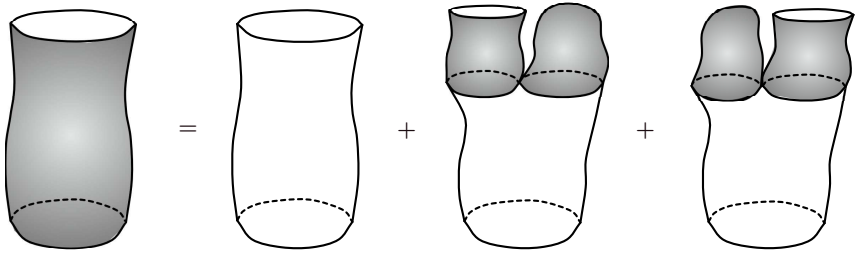}}}}
\caption{In all four graphs, 
the geodesic distance from the final to the initial 
loop is given by $t$. Differentiating
with respect to $t$ leads to eq.\ \rf{2.55}. Shaded parts of graphs represent
the full, $g_s$-dependent propagator and disc amplitude, and non-shaded 
parts the CDT propagator.}
\label{cap}
\end{figure}

In Fig.\ \ref{cap} we show what happens if we allow causality 
to be violated locally by allowing space to split in two at 
a certain time $t$, but we never allow the ``baby'' universe
which splits off to come back to the ``parent'' universe. 
The baby universe thus continues its life and is assumed
to vanish, shrink to nothing, at some later time. We now 
integrate over all such configurations in the path integral.
From the point of view of Euclidean space-time we are simply 
integrating over all space-times with the topology of a cylinder.
However, returning to the original Lorentzian picture it is clear
that at the point where space splits in two the light-cone is 
degenerate and one is violating causality in the strict local sense
that each space-time point should have a future and a past light-cone.
Similarly, when the baby universe ``ends'' its time evolution 
the light-cone structure is degenerate. These points thus have
a diffeomorphism invariant meaning in space-times with Lorentzian 
structure, and it makes sense to associated a coupling constant
$g_s$ with the process of space branching in two disconnected pieces.

The equation corresponding to Fig.\ \ref{cap} is \cite{alwz} 
\beq\label{2.55}
 \frac{\prt}{\prt t} \tG_{\lam,g_s}(x,y;t) = 
- \frac{\prt}{\prt x} \Big[\Big((x^2-\lam)+2 g_s\;\tW_{\lam,g_s}(x)\Big) 
\tG_{\lam,g_s}(x,y;t)\Big].
\eeq
$\tW_{\lam,g_s}(x)$ is denoted the disk amplitude with a fixed boundary cosmological
constant $x$. It is related to the disk 
amplitude with a fixed boundary length by
\beq\label{l-disk}
\tW_{\lam,g_s}(x) = \int_0^\infty \d l \; \e^{-x l} W_{\lam,g_s}(l).
\eeq
It describes the  ``propagation'' of the a spatial 
universe until it vanishes in the 
vacuum. If we did not allow any spatial branching we would 
simply have
\beq\label{disk}
\tW^{(0)}_\lam (x) = \int_0^\infty \d t \; G^{(0)}_\lam(x,l=0;t) = 
\frac{1}{x+\sqrt{\lam}},
\eeq
where $G^{(0)}_\lam(x,l; t)$ denotes the Laplace transform 
of $G^{(0)}_\lam(l',l; t)$ with respect to $l'$.
From the composition rules for $G_{\lam,g_s}(l_1,l_2;t)$ it 
follows that it has (mass) dimension 1. Thus $G_{\lam,g_s}(x,l_2;t)$ is 
dimensionless  and it follows that the (mass) dimension 
of the coupling constant $g_s$ must be 3. In a discretized theory
it will appear as the dimensionless  combination $g_s a^3$, $a$ being
the lattice spacing, and one can show that the creation 
of more than one baby universe at a given time $t$ is suppressed by 
powers of $a$ (see \cite{alwz} for details). 
Thus we only need to consider the process shown in Fig.\ \ref{cap}. 
For a fixed cosmological constant $\lam$ and boundary
cosmological constants $x,y$ expressions like  $\tG_{\lam,g_s}(x,y;t)$
and $\tW_{\lam,g_s}(x)$ will have a power series expansion in the dimensionless
variable
\beq\label{kappa}
\kp = \frac{g_s}{\lam^{3/2}}
\eeq
and the radius of convergence is of order one. Thus the coupling 
constant $g_s$ indeed acts to tame the creation of baby universes 
and if $g_s$ exceeds this critical value eq.\ \rf{2.55} breaks down and 
is replaced by another equation corresponding to Liouville quantum gravity
with central change $c=0$ (see \cite{alwz} for a detailed discussion).

\begin{figure}[t]
\centerline{\scalebox{0.5}{\rotatebox{0}{\includegraphics{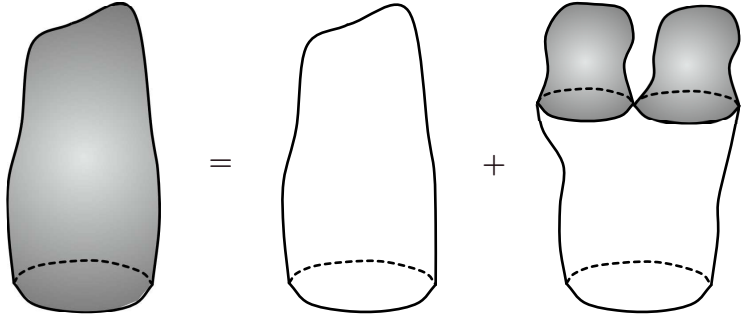}}}}
\caption[fig3]{{\small Graphical illustration of eq.\ \rf{3.2}. Shaded
parts represent the full disc amplitude, unshaded parts the CDT disc
amplitude and the CDT propagator. 
}}
\label{fig3}
\end{figure}

Differentiating the integral equation corresponding to Fig.\ \ref{cap}
with respect to the time $t$ one obtains \rf{2.55}.
The disc amplitude $\tW_{\lam,g_s}(x)$ is at this stage unknown. 
However, one has the
graphical representation for the disc amplitude 
shown in Fig.\ \ref{fig3}.
It translates into the equation \cite{alwz}
\bea\label{x3.2}
\lefteqn{\tW_{\lam,g_s} (x) = \tW_{\lam} ^{(0)}(x) +} \\
&&
g_s\int\limits_0^\infty \d t \int\limits_0^\infty \d l_1 \d l_2  \;
(l_1+l_2) G^{(0)}_{\lam}  (x,l_1+l_2;t) 
W_{\lam,g_s} (l_1)W_{\lam,g_s} (l_2) \nonumber
\eea
The superscript $(0)$  indicates the CDT amplitudes without 
baby universe branching, calculated above. We assume
\beq
\tW_{\lam,g_s=0} (x)=\tW^{(0)}_\lam(x),\label{relabel}
\eeq 
and similarly for $G^{(0)}_{\lam,g_s}$. 
The integrations in \rf{x3.2} can be performed and 
if we define $\hat{W}_{\lam,g_s}(x)$ by
\beq\label{x3.4}
\tW_{\lam,g_s}(x) = \frac{-(x^2-\lam) + \hat{W}_{\lam,g_s}(x)}{2g_s},
\eeq
one can show that 
$\hat{W}_{\lam,g_s}(x)$ is given by the expression
\beq\label{x3.9}
\hat{W}_{\lam,g_s}(x) = 
\lam(\tx-u)\sqrt{(\tx+u)^2-2\kp},
\eeq
where 
\beq\label{3.9a}
x= \tx\sla, ~~~~u^3-u+\kp=0.
\eeq
In order to have a physically acceptable $\tW_{\lam,g_s}(x)$, 
one has to choose the solution to the third-order 
equation which is closest to 1 and the above statements about the 
expansion of $\tW_{\lam,g_s}(x)$ in a power series in $\kp$ follow.

 Eq.\  \rf{2.55} can now be written as
\beq\label{x3.5}
 \frac{\prt}{\prt t} \tG_{\lam,g_s}(x,y;t) = 
- \frac{\prt}{\prt x} \Big[\hat{W}_{\lam,g_s}(x)\, \tG_{\lam,g_s}(x,y;t)\Big].
\eeq
In analogy with \rf{cdt32} and \rf{cdt33}, this is solved by
\beq\label{x3.6}
\tG_{\lam,g_s} (x,y;t) = 
\frac{\hat{W}_{\lam,g_s}(\bx(t,x))}{\hat{W}_{\lam,g_s}(x)} \; 
\frac{1}{\bx(t,x)+y},
\eeq
where $\bx(t,x)$ is the solution of the characteristic equation for \rf{x3.5},
the generalization of eq.\ \rf{cdt34}: 
\beq\label{x3.7}
\frac{\d \bx}{\d t} = -\hat{W}_{\lam,g_s}(\bx),~~~\bx(0,x)=x,
\eeq
such that
\beq\label{x3.8}
t = \int^x_{\bx(t)} \frac{\d y}{\hat{W}_{\lam,g_s}(y)}.
\eeq
This integral can be expressed in terms of elementary functions and one can 
thus find an explicit expression for $\tG_{\lam,g_s} (x,y;t)$ in the same way as 
eq.\ \rf{cdt35} led to an explicit solution for  the $\tG_{\lam}^{(0)} (x,y;t)$ appearing in 
\rf{cdt33} .

\section{The matrix model representation}\label{matrix}

The formulas \rf{x3.9} and \rf{x3.4} are standard
formulas for the resolvent of a Hermitean matrix model, calculated
to leading order in $N$, the size of the matrix. 
In fact  the following matrix model 
\beq\label{3.26}
Z(\lam,g_s) = \int d\phi \; 
\e^{-N\Tr V(\phi)},~~~~~
V(\phi) =  \frac{1}{g_s} \Big(\lam \phi -\frac{1}{3} \phi^3\Big)
\eeq
has a resolvent
\beq\label{resolv}
 \left\la \frac{1}{N}\;\Tr \left(\frac{1}{x-\phi}\right)\right\ra =
\tW_{\lam,g_s(x)}(x) + O(1/N^2),
\eeq
where $\tW_{\lam,g_s}(x)$ is given by \rf{x3.4}, and 
where the expectation value of a matrix expression $\cO(\phi)$ is 
defined as 
\beq\label{expect}
 \la \cO(\phi)\ra = 
\frac{1}{Z(\lam,g_s)}  \int d\phi \; 
\e^{-N\Tr V(\phi)} \; \cO(\phi).
\eeq

What is surpising here, compared to ``old'' matrix model approaches
to 2d Euclidean quantum gravity, is that the large $N$ limit reproduces
directly the continuum theory. No scaling limit has to be taken.
The situation is more like in the Kontsevich matrix model, which
directly describes continuum 2d gravity aspects. In fact the qubic 
potential is ``almost'' like the qubic potential in the Kontsevich 
matrix model, but the wold-sheet  interpretation is  different.

Can the above correspondance be made systematic in an 
large $N$ expansion and can the matrix model representation 
help us to a non-perturbative definition of generalized 2d
CDT gravity ? The answer is yes \cite{alwwz}.

First we have to formulate the CDT model {\it from first 
principles} such that 
we allow for baby universes to join the ``parent'' universe
again, i.e.\ we have to allow for 
topology changes of the 2d universe, and next we have to 
check if this generalization is correctly captured by the matrix 
model \rf{3.26} \cite{cdt-sft}.

\section{CDT string field theory}\label{sft}

In quantum field theory particles can be created and annihilated
if the process does not violate any conservation law of the
theory. In string field theories one operates in the 
same way with operators which can create and annihilate strings.
From the 2d quantum gravity point of view we thus have a 
third-quantization of gravity: one-dimensional universes can 
be created and destroyed. In \cite{sft} such a formalism was
developed for non-critical strings (or 2d Euclidean quantum 
gravity). In \cite{cdt-sft} the formalism was applied to
2d CDT gravity leading to a string field theory or third quantization 
for CDT, which allows us in principle to calculate any
amplitude involving creation and annihilation of universes.

Let us briefly review this formalism.
The starting point is the assumption of a vacuum from 
which universes can be created. We denote this state $\vac$ and
define creation and annihilation operators:
\beq\label{s1} 
[\Psi(l),\Psi^\dg(l')]=l\del(l-l'),~~~\Psi(l)\vac = \cav \Psi^\dg(l) =0. 
\eeq
The factor $l$ multiplying the delta-function is 
introduced for convenience, see \cite{cdt-sft} for a discussion.

Associated with the spatial universe we have a Hilbert space on the
positive half-line, and a corresponding scalar product (making $H_0(l)$
defined in eq.\ \rf{35b} hermitian) :
\beq\label{s4}
\la \psi_1 |\psi_2\ra = \int \frac{dl}{l} \; \psi_1^* (l) \psi_2(l).
\eeq
 
The introduction of the operators $\Psi(l)$ and $\Psi^\dg(l)$ in \rf{s1}
can be thought of as analogous to the standard second quantization 
in many-body theory. The single particle Hamiltonian $H_0$ defined by
\rf{35b} becomes in our case the ``single universe'' Hamiltonian. 
It has eigenfunctions
$\psi_n(l)$ with corresponding eigenvalues $e_n= 2n\sla$, $n=1,2,\ldots$:
\beq\label{s4a}
\psi_n(l) = l\, e^{-\sla l} p_{n-1}(l),~~~~~
H_0(l)\psi_n(l)= e_n \psi_n(l),
\eeq
where $p_{n-1}(l)$ is a polynomial of order $n\mi 1$.
Note that the disk amplitude $W^{(0)}_\lam(l)$, which is obtained from \rf{disk},
formally corresponds to $n=0$ in \rf{s4a}:
\beq\label{s4b}
W_\lam^{(0)}(l)=\e^{-\sla l},~~~~H_0(l) W_\lam^{(0)}(l)=0.
\eeq
This last equation can be viewed as a kind of Wheeler-deWitt equation
if we view the disk function as the Hartle-Hawking wave function. However,
$W^{(0)}_\lam(l)$ does not belong to the spectrum of $H_0(l)$ since it 
is not normalizable when one uses the measure \rf{s4}.

We now introduce creation and
annihilation operators $a_n^\dg$ and $a_n$ corresponding to these states,
acting on the Fock-vacuum $\vac$ and satisfying $[a_n,a^\dg_m]=\del_{n,m}$. 
We define
\beq\label{s5}
\Psi(l) = \sum_n a_n \psi_n(l),~~~~\Psi^\dg(l) = \sum_n a_n^\dg \psi^*_n(l),
\eeq
and from the orthonormality of the eigenfunctions with respect to 
the measure $dl/l$ we recover \rf{s1}. The ``second-quantized'' Hamiltonian is
\beq\label{s6}
\hH_0 = \int_0^\infty \dll \; \Psi^\dg (l) H_0(l) \Psi(l),
\eeq
and the propagator $G_\lam^{(0)} (l_1,l_2;t)$ is now obtained as
\beq\label{s7}
G_\lam^{(0)} (l_1,l_2;t) = \cav \Psi(l_2) \e^{-t \hH_0} \Psi^\dg(l_1) \vac.
\eeq 

\begin{figure}[t]
\centerline{\scalebox{0.45}{\rotatebox{0}{\includegraphics{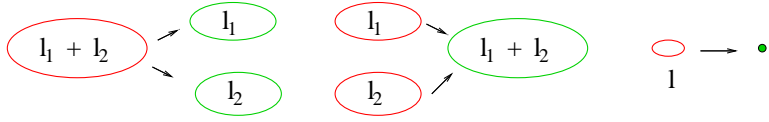}}}}
\caption[fig3]{{\small Graphical illustration of the various terms 
in eq.\ \rf{s8}.  
}}
\label{sft-fig}
\end{figure}

While this is  trivial, the advantage of the formalism
is that it automatically takes care of symmetry factors (like in the 
many-body applications in statistical field theory) both when many 
spatial universes are at play and when they are 
joining and splitting. We can follow
\cite{sft} and define the following Hamiltonian, describing the 
interaction between spatial universes:
\bea\label{s8}
\hH = \hH_0 &&-~ g_s 
\int dl_1 \int dl_2 \Psi^\dg(l_1)\Psi^\dg(l_2)\Psi(l_1+l_2)
\\ && - \a g_s\int dl_1 \int dl_2 \Psi^\dg(l_1+l_2)\Psi(l_2)\Psi(l_1)
-\int \dll \; \rho(l) \Psi(l), \nonumber
\eea
where the different terms of the Hamiltonian are illustrated in 
Fig.\ \ref{sft-fig}.
Here $g_s$ is the coupling constant we have already 
encountered in Sec.\ \ref{cap1}
of mass dimension 3. The factor $\a$ is just inserted to be 
able the identify the action of the two $g_s$-terms in \rf{s8} when
expanding in powers of $g_s$. We will think of $\a=1$ unless explicitly
stated differently.   
When $\a=1$ $\hH$ is hermitian except for the presence of the tadpole term.
It tells us that universes can vanish, but not be created from nothing.
The meaning of the two interaction terms is as follows: the first term
replaces a universe of length $l_1+l_2$ with two
universes of length $l_1$ and $l_2$. This is one of the 
processes shown in Fig.\ \ref{sft-fig}. The second term represents
the opposite process where two spatial universes merge into one,
i.e.\  the time-reversed picture. The coupling constant $g_s$ clearly 
appears as a kind of string coupling constant: one factor $g_s$ for 
splitting spatial universes, one factor $g_s$ for merging spatial
universes and thus a factor $g^2_s$ when the 
space-time topology changes, but there is also factors for 
branching alone. This is only compatible with an Euclidean SFT-picture
if we associate a puncture (and thus a topology change) with the
vanishing of a baby universe. As discussed above this is indeed not
unnatural from a Lorentzian point of view. From this point of view
the appearance of a tadpole term is more 
natural in the CDT framework than in the original Euclidean 
framework in \cite{sft}. The tadpole 
term is a formal realization of this puncture ``process'', where the 
light-cone becomes degenerate. 

In principle we can now calculate the process where we start
out with $m$ spatial universes at time 0 and end with $n$ universes
at time $t$, represented as
\beq\label{s100}
G_{\lam,g_s}(l_1,..,l_m;l'_1,..,l'_n;t) =
\cav \Psi(l'_1)\ldots \Psi(l'_n) \; 
e^{-t\hH}\Psi^\dg(l_1)\ldots \Psi^\dg(l_m)\vac.
\eeq

\subsection{Dyson-Schwinger equations}\label{ds}

The disk amplitude is one of a set of functions for which 
it is possible to derive Dyson-Schwinger equations (DSE).
The disk amplitude is characterized by the fact that at $t=0$ we
have a spatial universe of some length, and at some point it vanishes
in the "vacuum".   
Let us consider the more general situation where a set of spatial universes of some 
lengths $l_i$ exists at time $t=0$ and where the universes vanish at later times.
Define the generating function:
\beq\label{ds1}
Z(J)= \lim_{t\to \infty} \cav \e^{-t \hH} \; \e^{\int dl \, J(l) \Psi^\dg(l)}\vac.
\eeq
Notice that if the tadpole term had not been present in $\hH$ $Z(J)$ 
would trivially be equal 1. We have 
\beq\label{ds2}
\lim_{t \to \infty}  \cav \e^{-t \hH} \;\Psi^\dg(l_1)\cdots \Psi^\dg(l_n)\vac = 
\left.\frac{\del^n Z(J)}{\del J(l_1)\cdots \del J(l_n)}\right|_{J=0}.
\eeq
$Z(J)$ is the generating functional for universes that disappear in the
vacuum. We now have
\beq\label{ds3}
0= \lim_{t\to \infty}\left[\frac{\prt}{\prt t}\; 
\cav \e^{-t\hH} \; \e^{\int dl \, J(l) \Psi^\dg(l)}\vac =
-\cav \e^{-t\hH} \; 
\hH\;\e^{\int dl \, J(l) \Psi^\dg(l)}\vac\right].
\eeq
Commuting the $\Psi(l)$'s in $\hH$ past the source term effectively replaces
these operators by $l J(l)$, after which they can be moved to the left of 
any $\Psi^\dg (l)$ and outside  $\cav$. 
After that the remaining $\Psi^\dg(l)$'s in $\hH$ can
be replaced by $\del/\del J(l)$ and also moved outside
$\cav$, leaving us with a integro-differential operator acting on $Z(J)$:
\beq\label{ds4}
0= \int_0^\infty dl \, J(l) \,O \left(l,J,\frac{\del}{\del J}\right)Z(J)
\eeq
where
\bea\label{ds5}
\lefteqn{\hspace{-1cm}O \left(l,J,\frac{\del}{\del J}\right)= H_0(l) \frac{\del}{\del J(l)} -
\del (l)} \\
&& -g_s l \int_0^l dl'\frac{\del^2}{\del J(l')\del J(l-l')}
-\a g_s l\int_0^\infty dl' l'J(l')\frac{\del}{\del J(l+l')}\nonumber
\eea

$Z(J)$ is a generating functional 
which also includes totally disconnected universes which 
never ``interact'' with each other.  The generating 
functional for connected universes is obtained in the standard
way from field theory by taking the logarithm of $Z(J)$. Thus 
we write:
\beq\label{ds6}
F(J) = \log Z(J),
\eeq
and we have
\beq\label{ds7}
\lim_{t \to \infty} \cav \e^{-t\hH} \Psi^\dg(l_1)\cdots \Psi^\dg(l_n)\vac_{con} =
\left.\frac{\del^n F(J)}{\del J(l_1) \cdots \del J(l_n)}\right|_{J=0}, 
\eeq
and we can readily transfer the DSE \rf{ds4}-\rf{ds5} into an equation 
for the connected functional $F(J)$:
 \bea
0= \int_0^\infty dl \, J(l)
\left\{  H_0(l)\, \frac{\del F(J)}{\del J(l)} - \delta(l) 
 -g_s l \int_0^l dl'\;\frac{\del^2 F(J)}{\del J(l')\del J(l-l')} \right. 
\nonumber\\
\left. -g_s l \int_0^l dl'
\frac{\del F(J)}{\del J(l')}\frac{\del F(J)}{\del J(l-l')}
-\a g_s l\int_0^\infty dl' l'J(l')\frac{\del F(J)}{\del J(l+l')}\right\}.
\label{ds9}
\eea
From eq.\ \rf{ds9} one obtains the DSE by differentiating 
\rf{ds9} after $J(l)$ a number of times and then taking $J(l)=0$.

\subsection{Application of the DSE}\label{application}

Let us introduce the notation 
\beq\label{ds10}
w(l_1,\ldots,l_n) \equiv 
\left.\frac{\del^n F(J)}{\del J(l_1) \cdots \del J(l_n)}\right|_{J=0}
\eeq
as well as the Laplace transform $ \tw(x_1,\ldots,x_n)$.
Let us differentiate eq.\ \rf{ds9} after $J(l)$ one and two
times, then take $J(l)=0$
and Laplace transform the obtained equations. We obtain the 
following equations (where $H_0(x)f(x) = \prt_x [(x^2-\lam) f(x)]$):
\bea\label{ds13}
0&=&H_0(x)\tw(x) -1 +
g_s\prt_x \Big(  \tw(x,x) +  \tw(x)\tw(x)\Big),\\
&& ~ \nonumber\\
0&=&(H_0(x)+H_0(y))\tw(x,y) +g_s\prt_x \tw(x,x,y)+
g_s\prt_y  \tw(x,y,y) \label{ds15}\\
&& +2g_s\left(\prt_x [\tw(x)\tw(x,y)] \pl \prt_y[ \tw(y) \tw(x,y)]\right) 
+2\a g_s\prt_x\prt_y \Big(\frac{\tw(x)\mi \tw(y)}{x-y}\Big)\nonumber
\eea
The structure of the DSE for an increasing number of 
arguments is hopefully  clear (see \cite{cdt-sft} for details).

We can solve the DSE iteratively. For this purpose let us introduce 
the expansion of $\tw(x_1,\ldots,x_n)$
in terms of the coupling constants $g_s$ and $\a$:
\beq\label{ds12}
\tw(x_1,\ldots,x_n) = \sum_{k=n-1}^\infty \a^k\sum_{m=k-1}^\infty g_s^m \; 
\tw(x_1,\ldots,x_n;m,k).
\eeq
The amplitude $\tw(x_1,\ldots,x_n)$ starts with the 
power $(\a g_s)^{n-1}$ since we have to perform $n$ mergings 
during the time evolution in order to create a connected geometry
if we begin with $n$ separated spatial loops. Thus one can 
find the lowest order contribution to $\tw(x_1)$ from \rf{ds13}, use that
to find the lowest order contribution to $\tw(x_1,x_2)$ from \rf{ds15}, etc. 
Returning to eq.\ \rf{ds13}
we can use the lowest order expression for $\tw(x_1,x_2)$ to find the 
next order correction to $\tw(x_1)$, etc. 

As mentioned above the amplitude $\tw(x_1,\ldots,x_n)$ starts with the 
power $(\a g_s)^{n-1}$ coming from merging the
$n$ disconnected spatial universes. The rest of the powers
of $\a g_s$ will result in a topology change of the resulting, connected 
worldsheet. From an Euclidean point of view it is thus more appropriate
to reorganize the series as follows
\bea\label{ds12a}
&&\tw(x_1,\ldots,x_n) = (\a g_s)^{n-1}
\sum_{h=0}^\infty (\a g^2_s)^h \tw_h(x_1,\ldots,x_n)\\
&&\tw_h(x_1,\ldots,x_n) = \sum_{j=0}^\infty g_s^j 
 \tw(x_1,\ldots,x_n;n\mi 1\pl 2h \pl j,n\mi 1\pl h)\label{ds12b}
\eea
and aim for a topological expansion in $\a g^2_s$, at each order
solving for all possible baby-universe creations which at some 
point will vanish into the vacuum. Thus $\tw_h(x_1,\ldots,x_n)$ 
will be a function of $g_s$ although we do not write it explicitly.
The DSE allow us to obtain the topological expansion 
iteratively, much the same way we already did as a power expansion 
in $g_s$.

\section{The matrix model, once again}\label{matrix2}

Let us consider our $N\times N$ Hermitian matrix with the 
qubic potential \rf{3.26} and define the observable 
\beq\label{yy11}
\tW(x_1,\ldots,x_n)_d = N^{n-2} 
\left\la (\Tr \left(\frac{1}{x_1-M}\right)\cdots 
\left(\tr \frac{1}{x_1-M}\right)
\right\ra,
\eeq
where the subscript $d$ refers to the fact that the correlator
will contain disconnected parts. We denote the connected
part of the correlator by $\tW(x_1\ldots,x_n)$.
 It is standard matrix model 
technology to find the matrix model DSEs for  $\tW(x_1\ldots,x_n)$. 
We refer to \cite{davidloop,ajm,am,ackm} for details. 
{\it One obtains precisely the same set of 
coupled equations as  \rf{ds13}-\rf{ds15}  if we identify}:
\beq\label{N}
\a = \frac{1}{N^2},
\eeq
and the discussion surrounding the expansion \rf{ds12a} is 
nothing but the standard discussion of the large $N$ expansion  
of the multi-loop correlators (see for instance \cite{ackm} or 
the more recent papers \cite{eynard,ce,eo1}).
Thus we conclude that there is a perturbative agreement 
between the matrix model \rf{3.26} and the CDT SFT in the 
sense that perturbatively:
\beq\label{perturbative}
\tW(x_1,\ldots,x_n) = \tw(x_1,\ldots,x_n).
\eeq 
In practice the SFT {\it is}  only defined perturbatively, although
in principle we have available the string field Hamiltonian. However,
we can now use the matrix model to extract non-pertubative 
information. The identification of the matrix model 
and the CDT SFT DSEs were based on \rf{N}, but in the  SFT
we are interested in $\a = 1$, i.e.\ formally in $N=1$, in which 
case the matrix integrals reduce to ordinary integrals.
This means that we will consider the entire sum 
over topologies ``in one go":
\beq\label{3.1}
Z(g,\lam) = \int \d m \; 
\exp \left[ -\frac{1}{g_s} 
\left( \lam m - \frac{1}{3}\; m^3\right)\right],
\eeq 
while the observables \rf{yy11} can be written as 
\beq\label{3.2}
\tW_d(x_1,\ldots,x_n) = \frac{1}{Z(g_s,\lam)} 
\int \d m\; \frac{\exp \left[ -\frac{1}{g_s} 
\left( \lam m - \frac{1}{3}\; m^3\right)\right]}{(x_1-m)\cdots (x_n-m)}.
\eeq
These integrals should be understood as formal power series
in the dimensionless variable $\kp$ defined by eq.\ \rf{kappa}.
Any choice of an integration contour which makes the integral well 
defined and reproduces the formal power series is a potential
nonperturbative definition of these observables. However, different
contours might produce different nonperturbative contributions
(i.e.\ which cannot be expanded in powers of 
$t$), and there may even be nonperturbative contributions 
which are not captured by any choice of integration contour. 
As usual in such situations, additional
physics input is needed to fix these contributions.

To illustrate the point, 
let us start by evaluating the partition function given in 
\rf{3.1}. We have to decide on an integration path in the 
complex plane in order to define the integral. One possibility is to take a 
path along the negative 
axis and then along either the positive or the negative imaginary 
axis. The corresponding integrals are 
\beq\label{3.2a}
Z(g_s,\lam)= \sqrt{\lam}\; \kp^{1/3} F_{\pm} (\kp^{-2/3}),~~~
F_{\pm} (\kp^{-2/3}) =2
\pi \; e^{\pm i\pi/6}{\rm Ai}(\kp^{-2/3}\e^{\pm 2\pi i/3}),
\eeq
where Ai denotes the Airy function. Both $F_\pm$ 
have the same asymptotic expansion
in $\kp$, with positive coefficients. Had we chosen the integration path 
entirely along the imaginary axis we would have obtained ($2\pi i$ times)
${\rm Ai}(\kp^{-2/3})$, but this has an asymptotic expansion 
in $\kp$ with coefficients of oscillating sign, which is at odds with its
interpretation as a probability amplitude. In the notation of \cite{as} we have
\beq\label{3.2b}
F_{\pm}(z) = \pi \Big({\rm Bi}(z) \pm i {\rm Ai}(z)\Big),
\eeq 
from which one deduces immediately 
that the functions $F_{\pm}(\kp^{-2/3})$ are not real.
However, since ${\rm Bi}(\kp^{-2/3})$ grows like 
$e^{\frac{2}{3\kp}}$ for small $\kp$ while ${\rm Ai}(\kp^{-2/3})$ 
falls off like $e^{-\frac{2}{3\kp}}$, 
their imaginary parts are exponentially small 
in $1/\kp$ compared to the real part, and therefore do not contribute to
the asymptotic expansion in $\kp$.
An obvious way to {\it define} a partition 
function which is real and shares the
same asymptotic expansion is by symmetrization,
\beq
\oh (F_+ +F_-)\equiv \pi {\rm Bi}.
\eeq
The situation parallels the one encountered in the double scaling limit of the 
``old'' matrix model \cite{david-x}, and discussed in detail in \cite{marino},
but is less complicated. We will return to a 
discussion of this in Sec.\ \ref{hamiltonian}. 

Presently, let us collectively denote by $F(z)$ any of the functions 
$F_{\pm}(z)$ or $\pi {\rm Bi}(z)$, leading to the
tentative identification
\beq\label{3.3}
Z(g_s,\lam) = \sqrt{\lam}\; \kp^{1/3} \, 
F\Big(\kp^{-2/3}\Big),~~~~F''(z) = z F(z),
\eeq 
where we have included the differential equation 
satisfied by the Airy functions for
later reference. Remarkably, this partition function was also 
found in \cite{jk}, 
where a double scaling limit of so-called branched polymers 
were studied. It reflects that a significant part of the dynamics associated 
with the branching, 
as reflected in figs.\ \ref{cap} and \ref{fig3}, are indeed 
captured by a branched polymer model. It should be no surprise that 
this is possible. Branched polymers play an important role in non-critical 
string theory \cite{ad,ajt} and even in 
higher dimensional Euclidean quantum gravity
\cite{ajjk}. However, the 2d CDT has a much richer set 
of observables than those 
encountered in the theory of branched polymers, 
for instance  the  observables 
$\tW_d(x_1,\ldots,x_n)$, the calculation of which 
we now turn to.

Let us introduce the dimensionless variables
\beq\label{3.4}
x= \tx\,\sqrt{\lam},~~m= g_s^{1/3} \b,~~~~~
\tW_d(x_1,\ldots,x_n) = \lam^{-n/2} \tw_d(\tx_1,\ldots,\tx_n).
\eeq
Assuming $\tx_k > 0$, we can write
\beq\label{3.5}
\frac{1}{\tx-\kp^{1/3}\,\b}
= \int_0^{\infty} \d\a
\; \exp\left[-\left(\tx-\kp^{1/3}\b\right)\;\a\right].
\eeq
We can use this identity to re-express the pole terms 
in eq.\ \rf{3.2} to obtain the 
integral representation 
\beq\label{3.6}
\tw_d(\tx_1,\ldots,\tx_n) =  
\int_0^{\infty}\prod_{i=1}^n \d \a_i
\; \e^{-(\tx_1\a_1+\cdots +\tx_n\a_n)}\; 
\frac{F\Big(\kp^{-\frac{2}{3}}-
\kp^{\frac{1}{3}}\sum_{i=1}^n\a_i\Big)}{F\Big(\kp^{-\frac{2}{3}}\Big)}
\eeq
for the amplitude with dimensionless arguments.
By an inverse Laplace transformation we thus obtain:
\beq\label{3.7}
W_d(l_1,\ldots,l_n) = 
\frac{F(\kp^{-2/3}-\kp^{1/3}\sqrt{\lam}\,(l_1+\cdots+l_n))}{F(\kp^{-2/3})}.
\eeq
For the special case $n=1$ we find
\beq\label{3.8}
W(l) = \frac{F(\kp^{-2/3}-\kp^{1/3}\sqrt{\lam}\,l)}{F(\kp^{-2/3})}
\eeq
for the disc amplitude, together with the remarkable relation
\beq\label{3.8a}
W_d(l_1,\ldots,l_n)=W(l_1+\cdots+l_n).
\eeq
By Laplace transformation this formula implies the relation
\beq\label{3.8c}
\tW_d(x_1,\ldots,x_n)= 
\sum_{i=1}^n \frac{\tW(x_i)}{\prod_{j\ne i}^n (x_j-x_i)}. 
\eeq
 From $\tW_d(x_1,\ldots,x_n)$ we can construct the connected 
 multiloop functions $\tW(x_1,\ldots,x_n)$ using standard field theory. 
 Let us remark that the asymptotic
expansion in $\kp$ of $\tW(x_1,\ldots,x_n)$ of course agrees with that obtained
by recursively solving the CDT Dyson-Schwinger equations.

\section{Stochastic quantization}\label{stochastic}

It is a most remarkable fact that the above mentioned results can all 
be understood as a result of stochastic  quantization of {\it space}. In this
picture time becomes the {\it stochastic time} related with the branching 
of space into baby universes and the original CDT model described in 
Sec.\ \ref{cdt} becomes the classical limit where no stochastic 
processes are present \cite{sto-alwz}.

Recall the Langevin stochastic differential equation for a single variable
$x$ (see, for example, \cite{zinn1,chai}).
\beq\label{x2.1}
\dot{x}^{(\n)}(t) = - f\Big(x^{(\n)}(t)\Big) + \sOm \;\n(t),
\eeq
where the dot denotes differentiation with respect to stochastic time $t$,
$\n(t)$ is a Gaussian white-noise term of unit width 
and $f(x)$ a dissipative drift force:
\beq\label{x2.2}
f(x) =  \frac{\prt S(x)}{\prt x}
\eeq
 The noise term creates a probability distribution of $x(t)$, reflecting
the assumed stochastic nature of the noise term, with an associated 
probability distribution 
\beq\label{x2.3}
P(x,x_0;t) = \lla \del (x-x^{(\n)}(t;x_0))\rra_\n,
\eeq
where the expectation value refers to an average over the Gaussian noise.
$P(x,x_0;t)$ satisfies the Fokker-Planck equation
\beq\label{x2.5}
\frac{\prt P(x,x_0;t)}{\prt t} = 
 \frac{\prt}{\prt x}\left( \oh\Om \frac{\prt P(x,x_0;t) }{\prt x} + 
f(x) P(x,x_0;t)\right).
\eeq
This is an imaginary-time Schr\"{o}dinger equation, with
$\sOm$ playing a role similar to $\hbar$. It enables us to write
$P$ as a propagator for a Hamiltonian operator $\hH$,
\beq\label{x2.6}
P(x,x_0;t) = \la x | \e^{-t\hH}|x_0\ra,~~~\hH= \oh \Om \hp^2 +i \hp \, f(\hx),
\eeq
with initial condition $x(t=0) = x_0$, and $\hp = -i \prt_x$. It follows that
by defining
\beq\label{x2.6a}
\tG(x_0,x;t) \equiv \frac{\prt}{\prt x_0}\, P(x,x_0;t) 
\eeq
the function $\tG(x_0,x;t)$ satisfies the differential equation
\beq\label{x2.6b} 
\frac{\prt \tG(x_0,x;t)}{\prt t} = 
 \frac{\prt}{\prt x_0}\left( \oh \,\Om\, \frac{\prt \tG(x_0,x;t)}{\prt x_0} - 
f(x_0)\;\tG(x_0,x;t)\right).
\eeq

Omitting the noise term corresponds
to taking the limit $\Om \to 0$. One can then drop the functional average
over the noise in \rf{x2.3} to obtain
\beq\label{x2.21}
P_{cl}(x,x_0;t) = \del(x-x(t,x_0)),~~~~\tG_{cl}(x_0,x;t) = 
\frac{\prt}{\prt x_0} \del(x-x(t,x_0)).
\eeq
It is readily verified that these functions satisfy 
eqs.\ \rf{x2.5} and \rf{x2.6b} with $\Om =0$. Thus we 
have for $S(x) = -\lam x +x^3/3$:
\beq\label{x2.22} 
\frac{\prt \tG_{cl}(x_0,x;t)}{\prt t} = 
 \frac{\prt}{\prt x_0} \Big( (\lam - x_0^2 )\, \tG_{cl}(x_0,x;t)\Big).
\eeq  

Comparing now eqs.\ \rf{cdt32} and \rf{x2.22}, 
we see that we can formally re-interpret 
$\tG_\lam^{(0)}(x_0,x;t)$ -- an amplitude 
obtained by nonperturbatively quantizing Lorentzian
pure gravity in two dimensions -- 
as the ``{\it classical} probability'' $\tG_{cl}(x_0,x;t)$ corresponding
to the action $S(x) = -\lam x +x^3/3$ of a zero-dimensional system
in the context of stochastic quantization, only is the boundary condition
different, since in the case of CDT $x$ is not an ordinary 
real variable, but the cosmological constant. The correct boundary conditions
are thus the ones stated in eqs.\ \rf{cdt78}--\rf{cdt79}

Stochastic quantization of the system amounts to replacing
\beq\label{y2.5}
\tG_\lam^{(0)} (x_0,x;t) \to \tG(x_0,x;t),
\eeq
where $\tG(x_0,x;t)$ satisfies the differential equation
corresponding to eq.\ \rf{x2.6b}, namely,
\beq\label{y2.6}
\frac{\prt \tG(x_0,x;t)}{\prt t} = 
 \frac{\prt}{\prt x_0}\left( g_s \frac{\prt}{\prt x_0} + \lam-x_0^2\right) 
\tG(x_0,x;t).
\eeq
We have introduced the
parameter $g_s:=\Om/2$, which will allow us to reproduce 
the matrix model and SFT results reported above.   

 A neat geometric interpretation
of how stochastic quantization can capture topologically nontrivial
amplitudes has been given in \cite{sto-kawai}. Applied to the present
case, we can view 
the propagation in stochastic time $t$ for a given noise term $\n(t)$ 
as classical in the sense that solving the 
Langevin equation \rf{x2.1} for $x^{(\n)}(t)$ iteratively gives
precisely the tree diagrams with one external leg 
corresponding to the action $S(x)$ (and including the derivative
term $\dot x^{(\n)}(t)$), with 
the noise term acting as a source term. Performing the functional 
integration over the Gaussian noise term corresponds to integrating out the 
sources and creating loops, or, if we have several independent trees,
to merging these trees and creating diagrams with several external legs.
If the dynamics of the quantum states of the spatial universe 
takes place via the strictly causal CDT-propagator $\hG_0 = \e^{-t \hH_0}$, 
a single spatial universe of length $l$ 
cannot split into two spatial universes. Similarly, no two spatial universes are 
allowed to merge as a function of stochastic time.
However, introducing the noise term {\it and} subsequently
performing a functional
integration over it makes these processes possible. 
This explains how the stochastic quantization can automatically generate
the amplitudes which are introduced by hand
in a string field theory, be it of Euclidean character as described
in \cite{sto-kawai}, or within the framework of CDT.

What is new in the CDT string field theory considered
here is that we can use the corresponding 
stochastic field theory to solve the model.
since we arrive at 
closed equations valid to all orders in the genus expansion.
Let us translate equations \rf{y2.6} to $l$-space 
\beq\label{3.11}
 \frac{\prt G(l_0,l;t)}{\prt t} =-H(l_0)\, G(l_0,l;t),
\eeq
where the {\it extended} Hamiltonian
\beq\label{3.12}
H(l) = -l \frac{\prt^2}{\prt l^2} +\lam l - g_s l^2 = H_0(l)-g_s l^2
\eeq
now has an extra potential term coming from the inclusion of branching 
points compared to the Hamiltonian $H_0(l)$ defined in \rf{35b}. 
It is truely remarkable that all branching and joining is contained in 
this simple extra term. Formally  $H(l)$ is a well-defined Hermitian operator with respect to the measure \rf{s4} (we will discuss some subtleties in the 
next section). 

We can now write down the generalization of  Wheeler-deWitt equation
\rf{s4} for the disk amplitude 
\beq\label{3.11a}
H(l) W(l)=0.
\eeq
Contrary to $W_\lam^{(0)}(l)$ appearing in \rf{s4},  $W(l)$ contains
all branchings and all topology changes, and the solution is precisely 
\rf{3.8}! This justifies the choice $g_s=\Om/2$ mentioned above.
Recall that  $E=0$ does not belong to the spectrum of 
$H_0(l)$ since $W_0(l)$ is not integrable at zero with respect to the measure
\rf{s4}. Exactly the same is true for the extended Hamiltonian $H(l)$ and 
the corresponding Hartle-Hawking amplitude $W(l)$

We have also as a generalization of \rf{ham}  that
\beq\label{3.13}
G(l_0,l;t) = \la l | e^{-t H(l)}|l_0\ra
\eeq
describes the nonperturbative propagation of a spatial loop 
of length $l_0$ to a spatial loop of length $l$ in proper 
(or stochastic) time $t$, now including the summation over all genera.

\section{The extended Hamiltonian}\label{hamiltonian}

In order to analyze the spectrum of $H(l)$,
it is convenient to put the differential operator into standard form.
After a change of variables
\beq\label{4.3}
l= \oh z^2,~~~~~\psi(l) = \sqrt{z} \phi(z),
\eeq
the eigenvalue equation becomes
\beq\label{4.4}
H(z)\phi(z) = E \phi(z),~~~~H(z) = -\oh \frac{\d^2}{\d z^2} 
+\oh \lam z^2 + \frac{3}{8z^2}-\frac{g_s}{4} z^4.
\eeq 
This shows that the potential is unbounded from below, but 
such that the eigenvalue spectrum is still discrete: whenever
the potential is unbounded below with fall-off faster than $- z^2$, the spectrum is discrete, reflecting the fact that the classical escape time to infinity is finite
(see \cite{ak} for a detailed discussion relevant to the present situation).
For small $g_s$,  there is a large barrier of height $\lam^2/(2g_s)$ 
separating the unbounded region 
for $l > \lam/g_s$ from the region $0 \leq l \leq \lam/(2g_s)$ where the 
potential grows. This situation is perfectly suited to applying a standard WKB 
analysis. For energies less than $\lam^2/(2g_s)$, the eigenfunctions
\rf{s4a} of $H_0(l)$ will be good approximations to those of $\hH(l)$. 
However, when $l > \lam/g_s$ the exponential fall-off of $\psi_n^{(0)}(l)$
will be replaced by an oscillatory behaviour, with the wave function falling 
off only like $1/l^{1/4}$. The corresponding $\psi_n(l)$ is still
square-integrable since we have to use the measure \rf{s4}.
For energies larger than  $\lam^2/(2g_s)$, the solutions will be 
entirely oscillatory, but still square-integrable.

Thus a somewhat drastic change has occurred in the quantum behaviour of the one-dimensional universe as a consequence of allowing topology changes.
In the original, strictly causal quantum gravity model 
an eigenstate $\psi_n^{(0)}(l)$ of the spatial universe had an average size
of order $1/\sqrt{\lam}$.
However, allowing for branching and topology change, the average size of the universe is now infinite!

As discussed in \cite{ak}, Hamiltonians with unbounded potentials like 
\rf{4.4} have a one-parameter family of selfadjoint extensions and we 
still have to choose one of those such that the spectrum of $H(l)$ can be determined unambiguously.
One way of doing this is to appeal again to stochastic quantization, 
following the strategy used by Greensite and Halpern \cite{gh}, 
which was applied to the double-scaling limit of matrix models in \cite{ag,ag1,ak}.
The Hamiltonian \rf{x2.6} corresponding to the Fokker-Planck equation \rf{y2.6}, namely,
\beq\label{4.5}
H(x)\psi(x) = -g_s \frac{\d^2 \psi(x)}{\d x^2} +\frac{\d}{\d x}
\left(\frac{\d S(x)}{\d x}\, \psi(x) \right),~~~~ 
S(x) = \left(\frac{x^3}{3}-\lam\,x\right),
\eeq  
is not Hermitian if we view $x$ as an ordinary real 
variable and wave functions $\psi(x)$ as endowed with the standard scalar
product on the real line. However, by a similarity transformation one can transform $H(x)$
to a new operator 
\beq\label{4.6}
\tH(x) = \e^{-S(x)/2g_s}H(x) \, \e^{S(x)/2g_s};~~~
\tilde{\psi}(x) = \e^{-S(x)/2g_s}\psi(x),
\eeq 
which {\it is} Hermitian on $L^2(R,dx)$.
We have 
\beq\label{4.7}
\tH(x)= -g_s\frac{\d^2}{\d x^2} +
\left(\frac{1}{4g_s} \left(\frac{\d S(x)}{\d x}\right)^2+
\oh \frac{\d^2 S(x)}{\d x^2}\right),
\eeq
which after substitution of the explicit form of the action becomes
\beq\label{4.7a}
\tH(x)=-g_s \frac{\d^2}{\d x^2} +V(x),~~~~V(x)= \frac{1}{4g_s} (\lam -x^2)^2+ x.
\eeq
The fact that one can write 
\beq\label{4.8}
\tH(x)=  R^{\dg}R,~~~~
R=-\sqrt{g_s}\frac{\d}{\d x} +\frac{1}{2\sqrt{g_s}}\frac{\d S(x)}{\d x}
\eeq
implies that the spectrum of $\tH(x)$ is positive, discrete and 
unambiguous. We conclude that the formalism of stochastic quantization
has provided us with a nonperturbative definition of the CDT 
string field theory.

\subsection*{Acknowledgments}
JA, RL, WW and SZ acknowledge support by
ENRAGE (European Network on
Random Geometry), a Marie Curie Research Training Network, 
contract MRTN-CT-2004-005616, and 
RL acknowledges support by the Netherlands
Organisation for Scientific Research (NWO) under their VICI
program.
SZ thanks the Department of Statistics at Sao Paulo University for kind
hospitality and acknowledges financial support of the ISAC program,
Erasmus Mundus.

\end{document}